\documentstyle[epsfig,rotate]{aipproc}

\newcommand{\etal}{ {\it et al.}}
\newcommand{\msun}{M$_{\odot}$}
\newcommand{\degg}{$^{\circ}$}
\begin{document}
\setlength{\parskip}{0pt}
\title{Galactic Black Hole Binaries: Multifrequency Connections\footnote{Invited Review
Article for the 4th Compton Symposium, Williamsburg, VA, April, 1997}}
\author{S. N. Zhang$^*$$^{\star}$, I. F. Mirabel$^{\dagger}$, B. A. Harmon$^*$, 
R. A. Kroeger$^{\ddagger}$, L. F. Rodriguez$^{\sharp}$, R. M. Hjellming$^{\flat}$
and M. P. Rupen$^{\flat}$} 
\address{$^*$ES-84, Marshall Space Flight Center, Huntsville, AL 35812\\
$^{\star}$Universities Space Research Association\\
$^{\dagger}$CEA-CEN Saclay, Service D'Astrophysique, 91191 Gif-Sur-Yvette, Cedex, France\\
$^{\ddagger}$Naval Research Lab., MS 4151, 4555 Overlook Ave., Washington, DC 20375\\
$^{\sharp}$Instituto de Astronomia UNAM, Apdo Postal 70-264, DF04510 Mexico City, Mexico\\
$^{\flat}$National Radio Astron. Obs., P.O. Box O, 1003 Lopezville Rd., Socorro, NM 87801}

\lefthead{Zhang et al.: Black Holes}
\righthead{Zhang et al.: Black Holes}
\maketitle
\begin{abstract}
We review the recent multifrequency studies of galactic black hole binaries, aiming at
revealing the underlying emission processes and physical properties in these systems.
The optical and infrared
observations are important for determining their system parameters, such as the companion star
type, orbital period and separation, inclination angle and the black hole mass. The
radio observations are useful for studying high energy electron acceleration process, jet
formation and transport. X-ray observations can be used to probe the inner accretion disk 
region in
order to understand the fundamental physics of the accretion disk in the strongest
gravitational field and the properties of the black hole. Future higher sensitivity and 
better resolution instrumentation will be needed to answer the many fundamental questions that
have arisen.
\end{abstract}

\section*{Introduction}

Significant progress has been made in the study of the galactic black hole binaries since the 
launch of the Compton Gamma Ray Observatory in April 1991. It is now widely 
believed that some X-ray binary systems harbor a stellar black hole at the center of the 
accretion disk in each system. A variety of high energy spectra and light curves have 
been observed from many of them. Some of them also exhibited highly relativistic jets, 
which are found to be in correlation with the high energy radiation. Since the first mass 
determination of the assumed black hole in Cyg X-1 about 25 years 
ago\cite{bolton,webster-murdin}, the second property, i.e., the spins, of black holes have 
also been inferred recently\cite{zhang-cui-chen}, and are considered to be the missing 
link in the proposed unification scheme of all types of black hole 
binaries\cite{zhang-cui-chen}.

Several review articles on galactic black hole binaries now exist, including those in the same
proceedings. In this 
review article, we try to avoid any significant overlapping with other review articles,  
by focusing on (a) summarizing these new and important multiwavelength observations, and 
(b) exploring the physical connections
between observations made at different wavelengths to obtain more complete pictures of 
these systems. 

\section*{Basic high energy characteristics}

All of these galactic black hole binaries were originally discovered in either X-ray (1-10 
keV) or hard X-ray (20-300 keV) bands. Their X-ray or hard X-ray luminosity 
dominates their total electromagnetic radiation energy output. In this section we 
discuss briefly their basic high energy emission characteristics, in order to explore the 
multiwavelength aspects of these systems. 

\subsection*{High Energy Continuum Spectra}
\begin{figure}
\centerline{
\psfig{figure=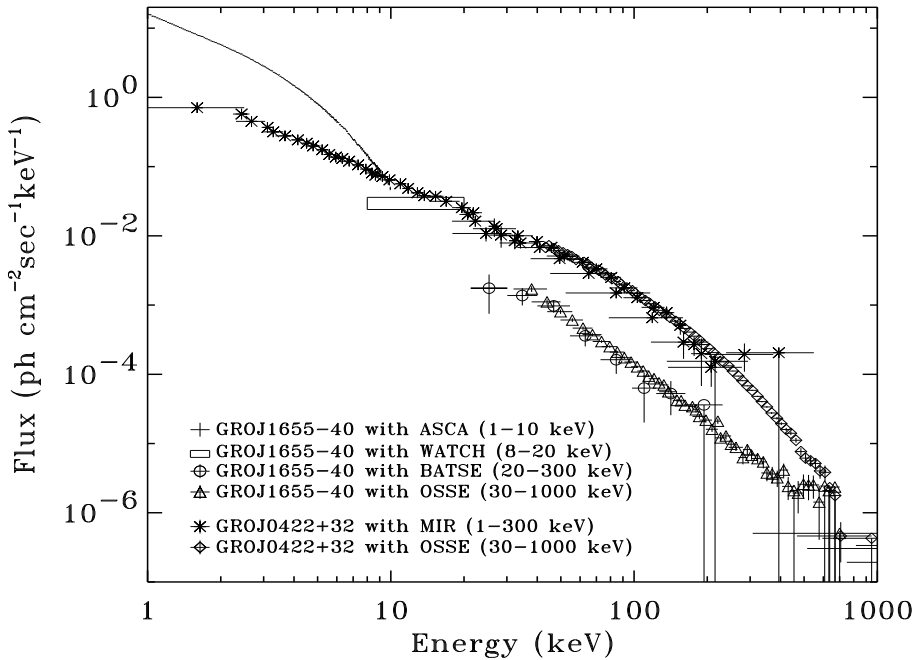,height=3.8in,width=5.0in}
}
\caption{Broad band high energy spectra of GROJ1655-40 [115] (Zhang \etal\ 1997) and GROJ0422+32 
[56] (Kroeger \etal\ 1996).}
\end{figure}
The high energy spectra of galactic (candidate) black hole binaries are made of primarily 
only two components -- a blackbody-like soft component and a power-law-like hard 
component (see \cite{zhang-pd} for a review of high energy continuum spectra observed in black
hole and neutron star X-ray binaries), as shown in figure 1 for the superluminal jet and black hole source 
GRO~J1655-40\cite{zhang-1655}. In some sources, the soft component is absent above 1 keV{\it and} the 
hard power-law component becomes harder with a break between 50-200 keV, as shown in the same figure 
for GROJ0422+32. The soft component is believed to be emitted from the 
inner accretion disk region and can be well described by the so-called 
multi-color disk blackbody spectral model\cite{makishima86,mitsuda84}, in which the total 
emitted spectrum is the
integration of the local blackbody emission from each annulus with a temperature 
$kT_{bb}\propto r^{-3/4}$. To obtain the physical parameters of the inner accretion 
disk, corrections for the relativity and spectral hardening effects must be taken into account
\cite{zhang-cui-chen}. The observed peak blackbody temperatures of the 
multi-color disk blackbody spectra vary between 0.1 to 2.5 keV (see figure 5 and the
corresponding section for details).

The origin of the power-law-like hard component is still not well understood, although 
inverse Compton up-scattering of low energy photons by high energy electrons are 
usually believed to be involved. The nature and origins of both the low energy photons 
and the high energy electrons are still
not identified unambiguously yet, despite the existence of many models. There usually 
exists an anti-correlation between the soft component luminosity and the hardness or 
flatness of the power-law whose photon spectral index varies between -1.5 to -3.5. 
Usually the power-law component is observed to fall into one of the two states: a hard 
state corresponding to a minimum luminosity 
($<5\times 10^{36}$ ergs/s)\cite{zhang-cygx-1} or the absence of soft X-ray 
component\cite{ebisawa-cha-ti}, with a photon spectral 
index between -1.5 to -2.0 and spectral break above 50-200 keV; or a soft state 
corresponding to a higher soft X-ray component 
luminosity\cite{ebisawa-cha-ti,grove-integral,zhang-1655,zhang-cygx-1}, with a photon 
spectral index between -2.5 to -3.5 and no detectable spectral break up to 300-600 
keV\cite{grove-integral}. Figure 1 illustrates a typical spectrum of each type (soft or high
state for GROJ1655-40 and hard or low state for GROJ0422+32). 
However, a source may remain in a different state for a 
time, and the transition or evolution between the states is usually continuous, for example 
in Cyg X-1\cite{zhang-cygx-1}. 

Both the soft and the hard components have been used as signatures or indicators of black 
hole binary systems. Similar components have, however, also been detected from neutron 
star binary systems, such as the type I X-ray burster 
4U1608-52\cite{mitsuda-1608,zhang-1608}, although the absolute hard X-ray luminosity is found
to be systematically higher in black hole systems than in neutron star systems
\cite{barret-ns-bh}. 

\subsection*{High Energy Long Term Light Curves}
\begin{figure}
\centerline{
\psfig{figure=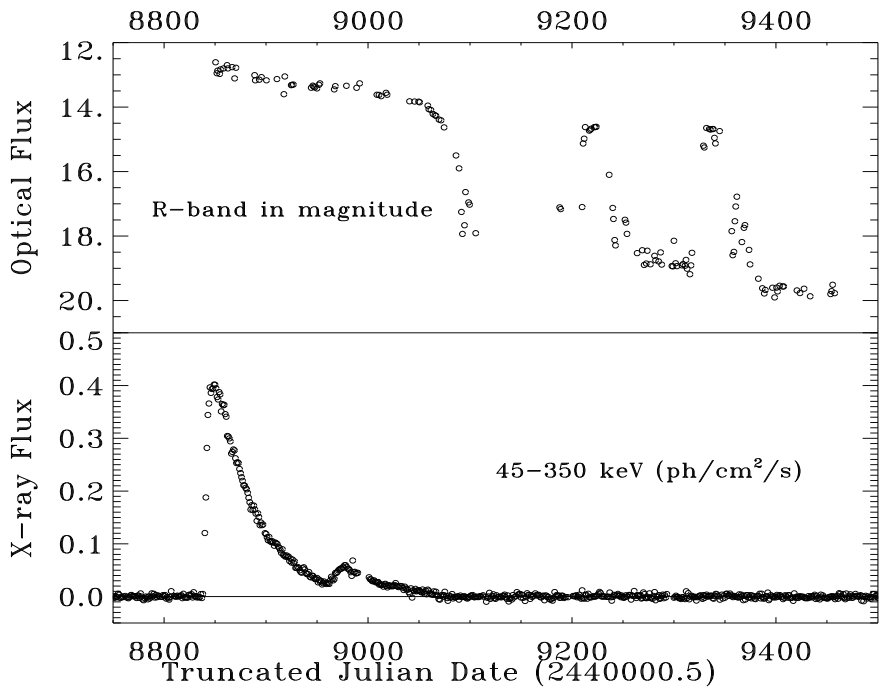,height=3.8in,width=5.0in}
}
\caption{Optical and hard X-ray light curves of a type I or nova-like source GROJ0422+32 
[11] (Callanan \etal\ 1996).}
\end{figure}

\begin{figure}
\psfig{figure=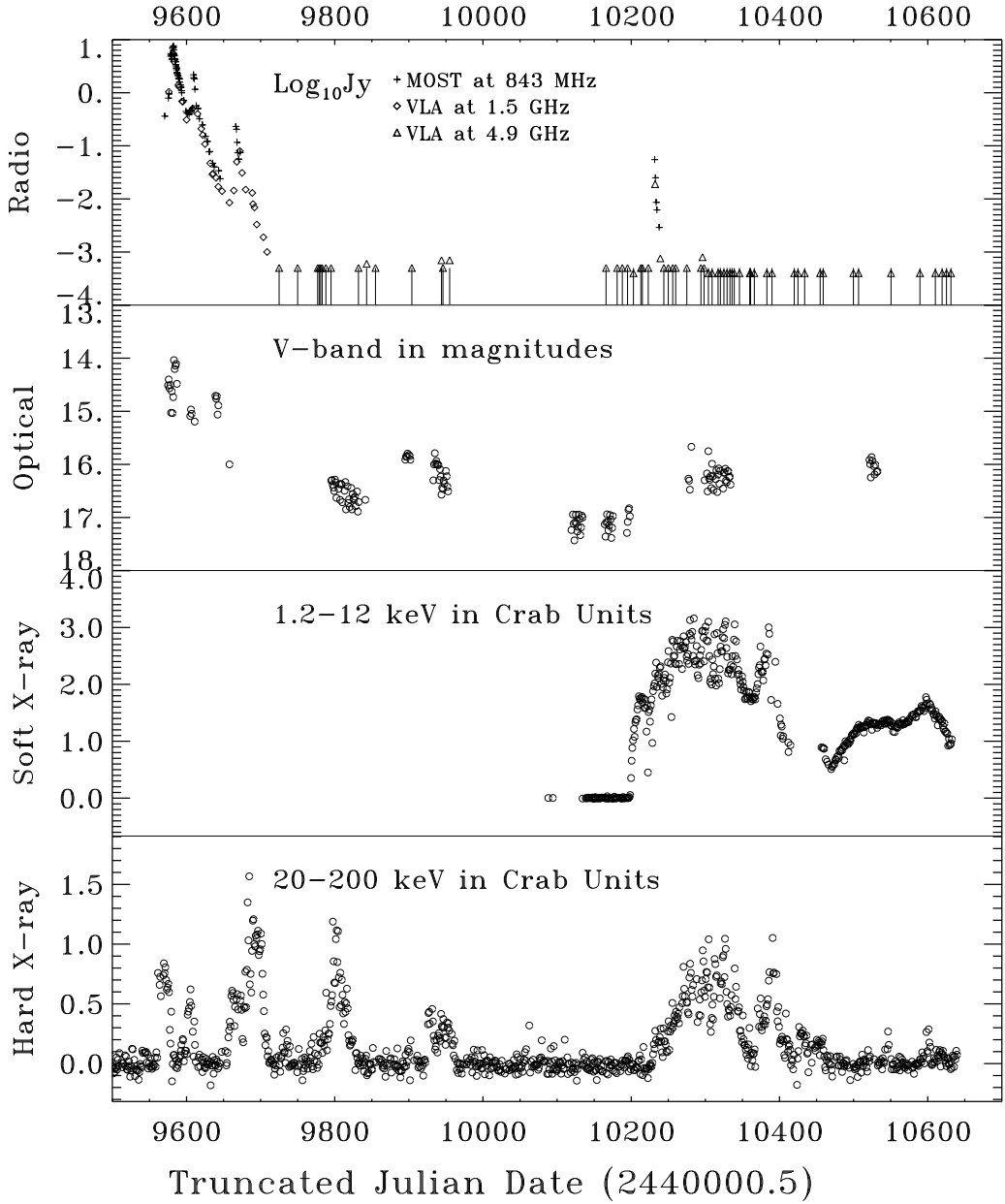,width=6.0in}
\caption{Radio [45,100,109] (Hjellming \& Rupen 1995, Tavani \etal\ 1996, Wu \& Hunstead 1996), 
Optical [2,3,76,78,80] (Bailyn \etal\ 1995a, 1995b, Orosz 1996, Orosz \& Bailyn 1997, Orosz \etal\
1997a, 1997b), soft and hard X-ray [90] (Robinson \etal\ 1997) light curves of a type II or 
multiple-outbursts source GROJ1655-40.}
\end{figure}

\begin{figure}
\psfig{figure=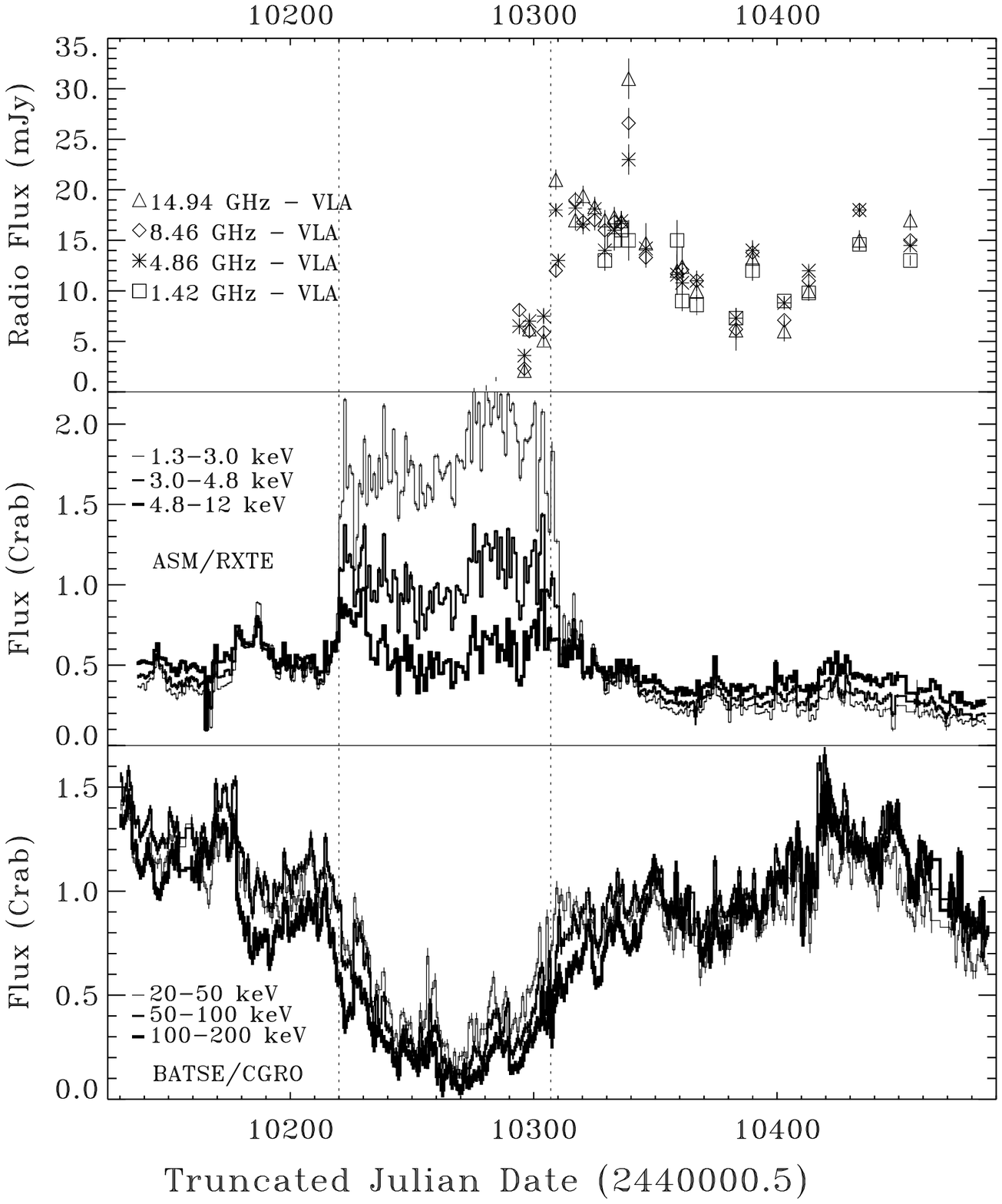,width=6.0in}
\caption{Radio, soft and hard X-ray light curves of the type III or persistent source Cygnus
X-1 [114,117] (Zhang \etal\ 1997).}
\end{figure}

A variety of high energy light curves have been observed from galactic black hole binaries
\cite{chen97}. These light curves can be placed conveniently into 
three classes: transients with nova-like light curves, which form the majority of the black hole 
binaries, such as A0620-00, GS2000+25, GS1124-68, GS2023+23, H1705-250, 
GROJ0422+32, GROJ1719-24 = GRS1716-249 and GRS1739-278, with a fast rise (a 
few days) and slow exponential decay (a few tens of days); transients with multiple 
outbursts, such as the superluminal jet sources GRS1915+105, GROJ1655-40 and a 
possible jet source GX339-4; persistent sources such as Cyg X-1, LMC X-1, LMC X-3. 
We shall call them type I or nova-like, type II or multiple-outbursts and type III 
or persistent {\it high energy}
light curves in the following discussion. Samples of some BATSE hard X-ray light curves, together with 
some optical, radio and soft X-ray light curves are shown 
in figures 2-4. Please refer to the references where some of the original light curves were
published for more details. (Note that some of the sources listed above, i.e., GROJ1719-24, GRS1739-278 and 
GX339-4, are not yet dynamically established black hole binaries with the compact object masses 
in excess of 3 \msun). These different types of light curves cannot be understood with high 
energy observations alone. In the following, we will try to understand some of their high 
energy behaviors using multifrequency data.

\begin{figure}
\psfig{figure=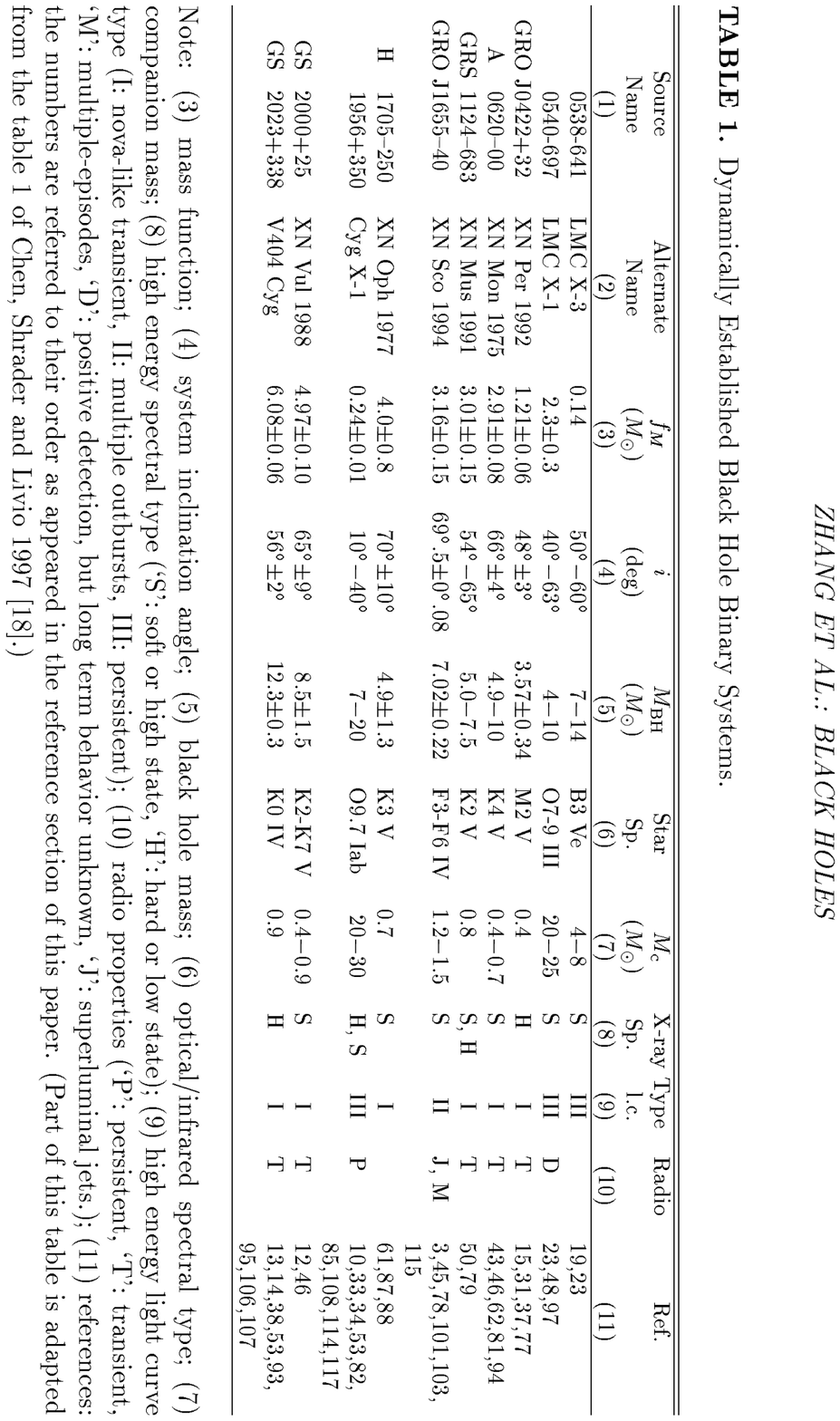}
\end{figure}

\section*{Optical and infrared studies}

In this section we discuss briefly the application of optical and infrared observations to 
the understanding of the properties of inner accretion disk regions and the black holes.

\subsection*{Determining the Black Hole Masses}

Ellipsoidal optical and infrared light curve modulations of the companions in these 
systems can be used to determine their orbital periods. Spectroscopic measurements of the 
Doppler shifts of some spectral line features can be used to obtain their mass functions, 
i.e., their mass lower limits. Detailed modelling of their light curves allows the 
determination of the system inclination angles. The spectral types of the companion 
provide information on the nature of the companion stars. Therefore in principle the 
complete system parameters, such as the orbital periods, orbital separations, masses of 
both the companions and the compact objects (the black hole for the systems we are 
discussing in this article) can be determined from optical and infrared
observations\cite{paradijs-clintock}. Perhaps the best example and 
success of this technique is the precise black hole mass (7 \msun) and other system 
parameters determination in 
GROJ1655-40\cite{hooft-1655-mass,orosz-1655-mass}.

Up to now, reliable dynamical mass determinations from optical and infrared observations 
have resulted in ten systems in which the compact object masses are most likely greater than 3.2 
\msun (see table 1, which lists also their radio and X-ray properties discussed in the
following sections), the observational and theoretical upper limits of a neutron star. In several systems, 
the lower limits to the compact object masses are already above 3.2 \msun. Arguably 
these constitute so far the strongest evidence of the existence of black holes in binary
systems. Although probing only the surface of the companion star and the outer disk 
region, these optical and infrared studies can also be used for understanding the high energy 
emission (from the inner-most region of the disk) behaviors of these systems, especially 
the three types of light curves we mentioned in the previous section.

\subsection*{Understanding Their High Energy Light Curves}

The companion stars in the systems producing the Type III or persistent light curves are all high mass 
(10-30 \msun) O or B stars. The type III light curves are apparently due to the high mass 
transfer rates  from their high mass companions, either through Roche-lobe overflow or 
stellar wind accretion. Apparently they are all accreting below the Eddington rate. There 
are, however, some significant differences between the three sources listed above. 
Cyg X-1 spends most of its time in a hard state, in which the total luminosity is dominated by 
the hard power-law. Occasionally the source enters into its soft state, in which a 
prominent soft component (with a higher temperature than that in the hard state) 
appears, nearly simultaneous with the softening of the power-law\cite{zhang-cygx-1}. LMC 
X-1 and LMC X-3, on the other hand, always remain in the soft state with prominent 
soft X-ray components and steeper
power-law components, similar to the soft state of Cyg X-1. Even between LMC X-1 and 
LMC X-3, there exists also some major differences. For example, in LMC X-1 the soft X-ray 
component remains rather stable, while the power-law component can vary 
substantially\cite{ebisawa89}. While in LMC X-3, the two components may vary 
substantially and independently, and the intensity and spectral hardness below 13 keV 
show a positive correlation\cite{ebisawa93}. It is also interesting to note that LMC X-3 
and Cyg X-1 are the only black hole binaries exhibiting long term 
periodic variations in their optical and X-ray light curves. In particular, orbital 
modulations in both the soft and hard X-ray bands have been seen from Cyg X-
1\cite{kitamoto,paciesas-4c,priedhorsky,zhang-cygx-1-5.6}. Although the soft X-ray flux 
modulation may be interpreted as  the modulation of the column density (ionized or 
neutral) due to stellar wind\cite{kitamoto}, the hard X-ray modulation is inconsistent with this 
interpretation\cite{paciesas-4c,priedhorsky95}. Since $\sim$ 30\% of the total hard X-ray component is 
estimated to be made of the reflection of an input power-law component \cite{done}, one possible 
origin of the observed orbital modulation is that its companion star also
contributes to the total reflection component. Then the contribution from the companion would
produce orbital modulations, similar to the optical and infrared ellipsoidal modulations. 

The companion stars in all these systems producing the Type I or nova-like light curves are low mass 
($<$1 \msun) 
main sequence stars, and no such low mass companion black hole binaries are known to 
produce other types of light curves. The transient nature of such systems has been 
explained as due to their sufficiently low mass transfer rate and that the mass accretor is a 
black hole\cite{KKS,paradijs96}. The outburst trigger mechanism is likely due to 
the thermal instability in the outer part of the 
disk\cite{lasota-iau}.   In the quiescence state, the material is steadily transferred at a 
very low rate from the companion to the accretion disk via Roche-lobe overflow and a 
significant portion of the material is accumulated in the disk. Thus the quiescence state disk 
is quasi-steady. When an instability 
is developed in the disk, an outburst occurs, during which the accumulated 
material is transferred to the black hole with a much higher rate. In this scenario the rise 
time would correspond to the sound propagation time from the outer disk to the inner 
disk\cite{mineshige96,rubin97}: $t_{rise}=R(\alpha C_{s})^{-1}$ (where $R$ is the outer disk 
radius, $\alpha$ the dimensionless viscosity parameter and $C_{s}$ the sound speed) 
which is on the order of a day for $\alpha \sim
0.1$. The decay time would correspond to the viscous time scale of the hot state disk,  
$t_{vis} = (R/H)^2(\alpha\Omega)^{-1}$ (where $H$ is the vertical height of the disk 
and $\Omega$ the angular speed of disk at radius $R$), which is on the order of 100 
days for $\alpha \sim 0.1$. 

There are still some fundamental questions 
concerning the quiescent disk structure. The standard thin disk extending to the last 
stable orbit of the black hole is problematic, because the inner part of the disk cannot 
stay in the `cold' branch of the `S' curve (inferred from the observed quiescence X-ray
emission), which is essential for the disk instability model\cite{lasota-iau}. The most recent 
version of the advection dominated
disk model\cite{narayan96-iau,hameury97} with the cooling dominated inner part truncated is much more successful 
than the standard thin disk model, but also has some problems. First, location of the inner
disk boundary, critical in the model, is not determined self-consistently, although the thermal
instability condition can be applied to produce a constrain. Second, it cannot explain the delayed hard X-ray outburst in 
GROJ1655-40\cite{robinson-4c}, with respect to the soft X-ray outburst after the initial optical 
on-set\cite{orosz97-optical}. Therefore a more consistent quiescent state disk model is still not 
available yet.

The companion stars in the Type II or multiple-outbursts systems may be intrinsically different from that 
in the Type I systems (the companion star in GX339-4 has not been identified yet). 
Although also called a 
low mass system, GROJ1655-40 has an evolved companion star of 2.3 
\msun\cite{bailyn95}. Usually such donor stars would have sufficiently high mass 
transfer rate, preventing the system from having a transient nature.  This discrepancy has been 
reconciled by assuming that GROJ1655-40 is in a short-lived evolutionary stage where 
the mass transfer rate is sufficiently low to allow instabilities to occur\cite{kolb97}. This
is a possible way to explain the rarity of this type of systems. 
Its multiple 
outbursts are not periodic and so apparently not related to its orbital motion, but most 
likely due to another instability after the initial outburst is 
triggered. Perhaps the X-ray heating of the companion star induces the so-called 
mass 
transfer instability \cite{chen-livio-gehrels,hameury86} or `echoes' \cite{augusteijn}. 
This may be due to its peculiar type (much massive and larger than other companion stars)
of this companion star. 
Comparing the the K(2.2$\mu$m) spectra of GRS1915+105 
observed at different epochs \cite{castro-tirado,fender-pri,mirabel-1915-ir}, 
one finds the characteristic 
HeI and Br$\gamma$ emission lines of O-Be stars \cite{mirabel-1915-ir} with time variable 
intensity. It has been shown that the absolute infrared 
magnitude, time variability, and spectral shape of GRS1915+105 is comparable 
to that of SS 433 and other high mass X-ray binaries \cite{chaty-96}. Furthermore, the infrared afterglow of an 
X-ray/radio outburst observed on August 1995 revealed that at that time GRS1915+105 was 
enshrouded in a dusty nebula of $\sim$ 10$^{16}$ cm radius\cite{mirabel-1915-96}. 
In summary, the infrared observations suggest that the donor star belongs to 
the class of massive stars with transitional spectral classification due to 
the dynamically unstable stellar atmosphere or wind \cite{morris}. 
If this is confirmed, GRS 1915+105 would be a black hole with a peculiar massive 
companion which is losing mass at rates of 10$^{-6}$-10$^{-5}$ M$_{\odot}$ 
yr$^{-1}$.
Further more definitive observations are still needed to clarify its companion star 
type, which may be critical for understanding the nature of the system. Despite that 
GROJ1655-40 is the only one of the Type II sources whose companion is 
unambiguously identified to be a peculiar star, it is worthwhile to explore the relationship 
between the properties of the companions and the high energy behaviors of the systems.

\section*{Radio studies}

Radio fluxes from these systems originate in the incoherent synchrotron radiation of high energy 
electrons. The radio emission region is usually comparable or larger than the size of 
the accretion disk or the whole binary system. 
For a typical source of 0.1 Jy at 3 kpc, since the surface brightness temperature cannot 
exceed 10$^{12}$ K, the minimum size of the radio emission region is about 10$^{12}$ 
cm. The size of a typical black hole binary with an orbital period of order one day is 
also about 10$^{12}$ cm. For those spatially resolved systems, i.e., those jet sources, their 
sizes are usually between 10$^{14}$ to 10$^{16}$ cm.  Therefore the radio emission 
region can be significantly larger than the whole binary system. At a distance far away 
from the accretion disk, the strength of the magnetic field is most likely of order
10$^{-3}$ Gauss, therefore the typical electron energy is of order 10$^{9}$ eV for a characteristic 
radiation frequency of 10$^{10}$ Hz. Therefore the radio emission from black hole binary 
systems is usually produced by high energy electrons far away from the central region of 
the accretion disk.

Corresponding to the three types of X-ray light curves, there are also three types of radio 
light curves, with similar morphologies. The systems producing the type I or nova-like X-ray light 
curves produce usually also type I radio light curves with fast rises (although their 
radio flux on-sets are rarely caught) and slow exponential decays, in good correlation with 
the X-ray light curves (see ref.\cite{hjellming-radio-review} for a review). The radio 
emission regions in this class of systems are never spatially resolved and have been 
modelled as synchrotron bubble events, i.e., expanding spherical bubbles of relativistic 
plasma. The origin and acceleration mechanism of the high energy electrons is still not 
identified, but is believed to be related to the enhanced mass accretion rate in the inner 
accretion disk region, inferred from the correlation between the radio and X-ray light 
curves. It is worthwhile to note that such mass accretion and radio emission correlation cannot be firmly 
established in  those active galactic nuclei (AGN), since the dynamical time scales in AGN 
are typically larger by a factor of 10$^{6-9}$ than the black hole binary systems.

The radio light curve of Cyg X-1 is very similar to its X-ray light curve, i.e., persistent, 
variable and with two distinct states. Its high radio flux state corresponds to its hard 
X-ray high and soft 
X-ray low flux state, i.e., the so-called hard/low state
\cite{zhang-cygx-1,zhang-cygx-1-radio}. Although the hard/low state mass accretion rate 
may be slightly lower than that in its soft/high state,  any changes in mass accretion rate 
through the inner accretion disk region may not play a strong role\cite{zhang-cygx-1}. Therefore the 
observed radio, soft and hard X-ray correlation suggests that the radio flux increase is not 
simply related to the mass accretion rate increase. Instead, the radio flux seems to respond 
to the hard X-ray flux positively. It is therefore possible that the much higher energy 
(10$^{9}$ eV) electrons responsible for the radio flux production are somehow related 
to the much lower energy (10$^{5}$ eV) electrons, despite that the former population 
of electrons are distributed far away from the central region of the disk, where the 
hard X-ray photons are produced. Perhaps the much higher energy electrons originate from the 
lower energy 
electrons and they share the same initial acceleration mechanism. 
Exactly how these electrons are accelerated and why the acceleration process can be 
maintained stably still remain poorly understood. Another type-III X-ray 
light curve source, LMC X-3, has never been observed  at any radio 
frequencies, with an upper limit of $<$ 0.3 mJy at 3 and 6 cm \cite{fender-pri}. 
This would not be 
surprising if its radio luminosity is similar to that from 
Cyg X-1, since it is located at about a factor of 20 farther away than Cyg X-1 and the 
radio flux from Cyg X-1 is only between a few and tens of mJy. 
The detection of a strong radio flux has been reported from the third type III source LMC X-1 
at 81 mJy at 6 
cm \cite{spencer97}. Assuming isotropy, its absolute luminosity is about a factor of 3000 
{\it stronger} than 
Cyg X-1, about a factor of 2 {\it brighter} than the peak luminosity of the superluminal jet sources 
GROJ1655-40, and {\it comparable} to the peak luminosity of the other superluminal jet 
source GRS1915+105. We caution the readers that a later observation of LMC X-1 with the same
instrument did not detect the source at an upper limit of $\sim$ 1 mJy \cite{fender-pri}.
Therefore more radio observations of the source are still needed to clarify the situation.

The radio light curves of the two superluminal jet sources, i.e., GROJ1655-40
\cite{hjellming-1655-nature,tingay} and 
GRS1915+105 \cite{mirabel-dis}, exhibit close, but complicated relationships with their type 
II X-ray light 
curves. Their peak radio luminosity is also much higher than most of the type I or III systems
discussed above. In GRS1915+105, there seems to exist an almost one to one correspondence 
between the radio flares and the {\it hard} 
X-ray outbursts\cite{foster-1915}. However, there seems to be three different kinds of 
radio flare states\cite{foster-1915}.  The relationship between the 
particular type of the radio flare and the hard X-ray emission properties is not clear. The 
correlation between radio intensity and the hard X-ray intensity is positive below about 100
mJy; above this level, the correlation becomes negative\cite{harmon-1915}.
In GROJ1655-40, a dynamically confirmed black hole binary, the initial hard X-
ray outburst\cite{zhang-1655-ini} was accompanied with a bright radio flare resulting from 
superluminal jet ejection\cite{hjellming-1655-nature,tingay}. All first three hard X-ray 
outbursts were well correlated with radio flares (superluminal ejection 
events)\cite{harmon-1655-nature}, although the overall level of the radio emission 
followed a slow exponential decay. Two subsequent hard X-ray outbursts, however, were 
not accompanied with any detectable radio emission above 5 mJy between 1.5-15 
GHz\cite{tavani-1655}. Therefore, this seems to indicate that significant hard X-ray 
production is a necessary, but not sufficient condition for jet ejection. Therefore 
models\cite{levinson-1655,meier-1655} involving hard X-ray production from the 
relativistic jets seem unlikely\cite{zhang-1655}. 
Nevertheless, what we have learned from 
the radio and hard X-ray correlations are still important for understanding some 
problems. 

The positive X-ray and radio flux correlation indicates that jet ejection events 
are related to increased mass accretion. It is widely accepted that a magnetic field should be 
an essential ingredient in the jet formation according to the Blandford and Payne 
model\cite{blandford-model,blandford-payne}. Therefore it is very likely that rather strong magnetic fields 
exist in these systems. If the hard X-ray component is indeed 
produced by the long suspected Comptonization in an optically thin (inferred from the 
rather steep power law) corona, then the correlation also supports the jet production 
model in which the jets are produced from the interaction between the corona and the 
magnetic field\cite{meier-corona}. 

The black hole (candidate) binary system GX339-4 is also a variable radio 
source\cite{sood-339}. Intense ($\sim$ Jy) radio flares similar to those from GROJ1655-40 and 
GRS1915+105 have never been observed. We, however, cannot conclude that the source 
has never produced radio flares because of the rather sporadic observations from the 
southern hemisphere. Because of this, the relationship between its radio and X-ray fluxes 
is not completely clear. The recent observation of a possible radio jet in GX339-4
\cite{fender-339}, during an X-ray active state, may provide evidence that its radio 
behavior is similar to the two superluminal jet sources. We may eventually be able 
to conclude that all the type II or multiple-outbursts systems share some common radio 
emission properties, related to their high energy, especially their hard X-ray, behavior.

An essential question to be answered concerns the physical connections between the 
types of the companion stars, orbital separations, optical and IR light curves, X-ray and 
hard X-ray light curves, and radio light curves. A possible scenario follows:  their 
companion stars and orbital separations determine the mass transfer property from the 
companion stars to the outer regions of the disk, this in turn controls their optical
properties, at 
least during the initial outburst after a quiescence period. The mass flow rates through the 
inner disk boundary onto the black hole  modulate their soft and hard X-ray light curves. 
The soft X-ray photons are emitted from the optically thick inner accretion disk region, 
and the hard X-ray photons are believed to be produced by inverse Compton up-scattering 
of low energy photons (probably the soft photons from the disk) by higher 
energy electrons, although the origin of the electrons is still not unambiguously identified. 
Somehow at least a portion of these electrons are accelerated to higher energies and move far away from 
the central disk region to produce radio fluxes via incoherent synchrotron radiation. This 
global picture agrees qualitatively to the three types of X-ray and radio light curves, 
through our investigations of the multifrequency connections in these systems. The 
correlation between the three types of companion stars and their corresponding
light curves suggest that in the type II or multiple-outbursts systems another instability, possibly the mass 
overflow instability in the companion star caused by the X-ray heating of the central soft 
and hard X-ray sources, may play a major role in the subsequent outbursts since the 
initial one. In the type III or persistent systems, perhaps the feedback of the X-ray heating to the 
high mass companion stars reaches a steady state so that these systems become persistent 
sources.

The above picture does not explain why some systems produce powerful jets, while 
others apparently do not do so. It is possible that in GROJ1655-40 the inferred mass 
transfer rate from the X-ray observations is only a portion of the total mass transfer rate 
from its peculiar companion star. The rest of the material is not transferred through the 
disk, but instead forms a disk `wind', available for producing the powerful jets. In other 
systems the disk `wind' may be very weak. In GRS1915+105, although the exact type of its 
companion is unknown as we 
discussed above, there indeed appears to be some evidence for a significant wind from the 
system\cite{mirabel-1915-ir}. This scenario, however, still does not explain why the 
outflows in GROJ1655-40 and GRS1915+105 are highly collimated. This may be 
partially explained by the properties of the black holes, as will be discussed below.

\section*{X-ray Probing of the Inner Disk Region}

X-ray observations at keV energies can be used for probing the accretion disk very close 
to the black hole horizon, thus providing important tools for testing accretion disk models 
and some general relativistic effects. This is because the peak emission energy of the 
blackbody spectra from the inner disk region is between one-tenth and several 
keV\cite{zhang-cui-chen}, and that these X-ray photons are much more penetrating than 
EUV photons emitting from the inner disk regions of AGNs. Because of the interactions 
between the black holes and the accretion disks, X-ray observations can also be used for 
studying the properties of the black holes\cite{zhang-cui-chen}. In this section we will 
discuss three aspects of probing the accretion inner disk regions with X-ray observations.

\subsection*{Continuum X-ray Spectra}
\begin{figure}
\centerline{
\psfig{figure=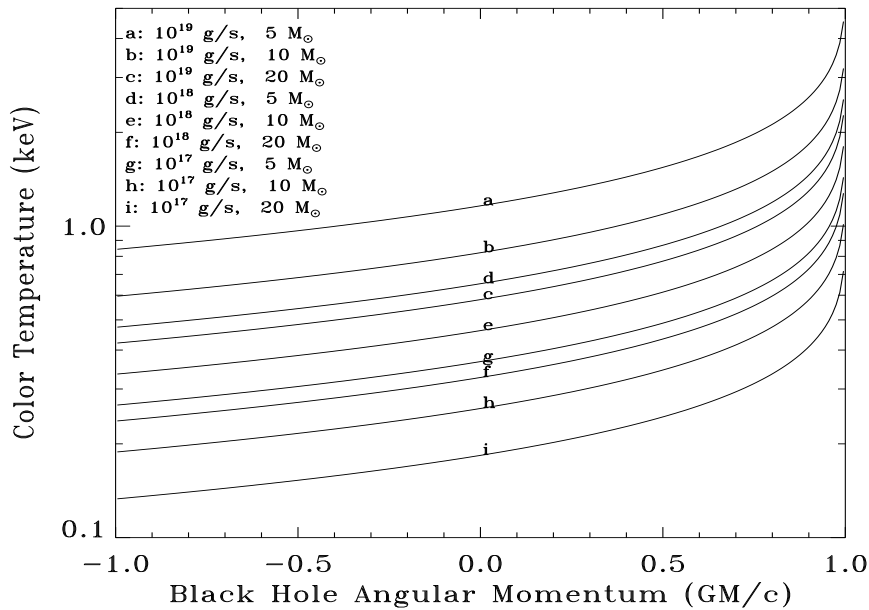,height=3.5in,width=5.0in}
}
\caption{The peak color temperature ($kT_{\rm col}$) of the accretion disk emission,
for different black hole masses, accretion rates and angular momenta
[116] (Zhang, Cui \& Chen 1997).}
\end{figure}

As we mentioned before, the continuum X-ray spectrum from the inner accretion disk 
region is well described by the multi-color disk blackbody 
model\cite{makishima86,mitsuda84}. The spectrum depends only upon the location of the 
inner accretion boundary, mass of the black hole and the mass flow rate through the inner 
disk region. For a steady disk in a high mass accretion rate state, the inner disk boundary is 
assumed to be the last stable orbit of the black hole, beyond which the material falls into 
the black hole with near radial trajectories. For the purpose of the spectral fitting, the free 
parameters are the peak blackbody temperature, absorption column density, and the 
normalization at a given energy. The inner disk radius and the disk inclination angle are 
coupled in the normalization parameter, so they cannot be determined independently. In some 
cases the system inclination angle and/or the black hole mass can be well constrained 
through the optical light curve modelling and spectroscopic measurements (to determine 
the companion star type). Therefore the size of the inner disk boundary can also be 
constrained. The values of the black hole mass, spin and the rotation direction of the 
material moving around it are all coupled into the size of the inner disk boundary, since the 
location of the last stable orbit of a black hole depends upon all of them. If the mass of 
the black hole is also known from optical observations, we then have the possibility to 
constrain the spin of the black hole. It should be noted that  
various correction factors to account for spectral hardening (due to electron scattering) 
and relativity effects have to be included in this process. This has been done recently in a 
number of black hole systems, whose system parameters have been well constrained from 
optical observations\cite{zhang-cui-chen}. In figure 5, the peak blackbody temperature is shown as the 
functions of the black hole mass, spin and mass accretion rate.

It has been found that the two superluminal jet sources GROJ1655-40 and GRS1915+105 (by
combining the interpretation of its 67 Hz QPOs as the $g$-mode oscillations\cite{nowak96})
contain
most likely a black hole spinning at near the maximally allowed rate, while several other
black holes with the observed soft components, but no relativistic jets, contains 
slowly or non-spinning black holes. It is also proposed that several `hard' X-ray transients, i.e.,
no soft component has ever been observed from them, may contain rapidly spinning black holes with 
retrograde
disks \cite{zhang-cui-chen}. Therefore all types of observed black hole binary systems are naturally unified 
within
one simple scheme. The state transitions in Cyg X-1 are proposed to be due to the disk
rotation direction reversal, caused by the so-called `flip-flop' type instability in wind
accreting systems\cite{zhang-cui-chen}. Future higher sensitivity observations, especially below 1 keV, are 
needed
to test this scheme.

\subsection*{Iron Line Diagnostics}

The first spectral line feature from a black hole binary system was detected with the {\it 
EXOSAT GSPC} in Cyg X-1 \cite{bar85}. Since then the Cyg X-1 spectral line feature has been 
observed with {\it Tenma}\cite{kitamoto90}, {\it Ginga}\cite{ebisawa92,tanaka91} and more 
recently with {\it ASCA}\cite{ebisawa96}. From the high resolution ASCA observations
\cite{ebisawa96}, the overall feature can be modelled by a narrow
(equivalent width 10-30 eV) iron {\it K} emission line at 6.4 keV and an iron $K$ edge at 7 keV
representing a reflection component, from the outer part of the accretion disk. The disk is inferred to be 
ionized,
with a covering angle $\Omega/2\pi \sim 0.2-0.4$. Due to the limited statistics, minor
contribution due to the reflections from the companion star and/or fluorescence emission from
the inner accretion disk
region cannot be excluded. The small or null contribution from the inner disk region may be due to
the much higher degree of ionization of the inner disk region. However, it should be noted that 
these observations were made during
the regular hard/low state of Cyg X-1. In the hard/low state, its inner disk radius 
is probably larger than 3$R_{s}$\cite{zhang-cygx-1}.
Therefore the inner disk contribution to the observed line profile is expected to be less
important in the hard/low state. Moreover, the broadening and skewness of the expected line
profile from the inner disk region is less significant for a small inclination angle (only
10\degg -40\degg\ in Cyg X-1). Future higher sensitivity and better resolution instruments are required
for unambiguously identifying these different components in its iron line profile.

\begin{figure}
\psfig{figure=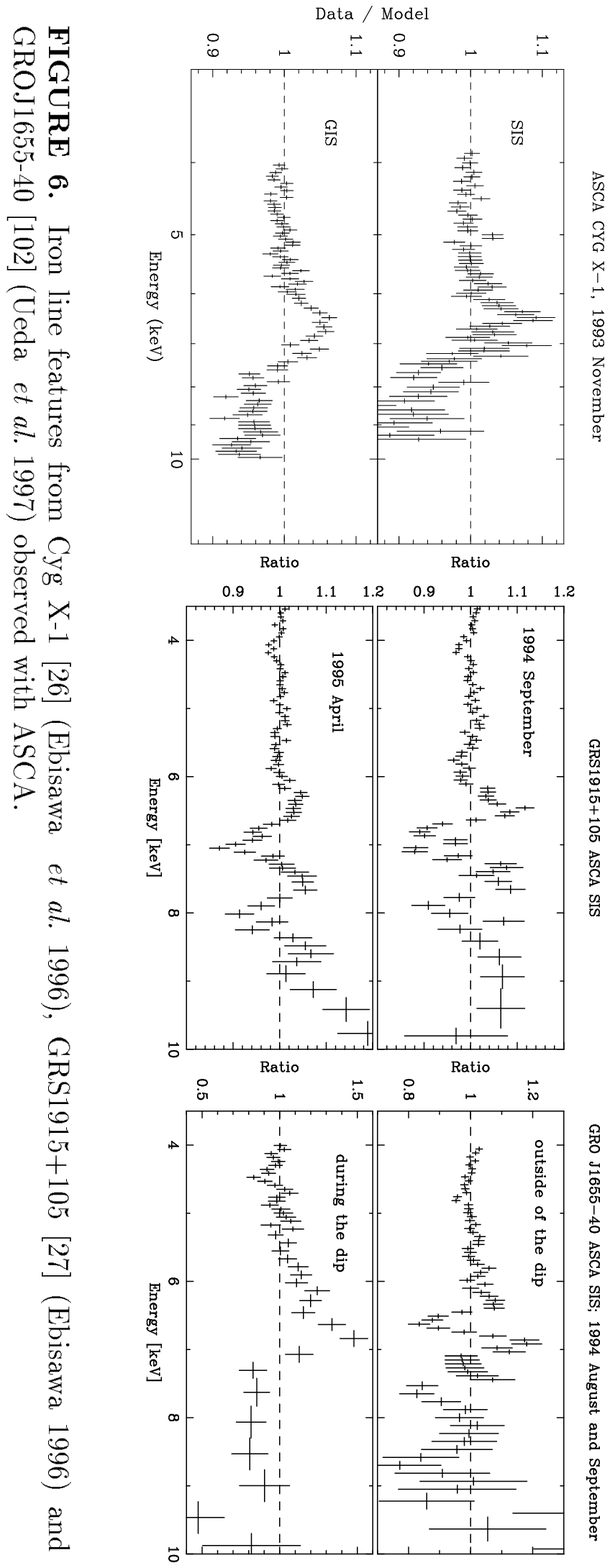}
\end{figure}

Recently, iron line features have also been observed from the two superluminal jet sources
GROJ1655-40 and GRS1915+105 with
ASCA\cite{ebisawa-asca}, as shown in figure 6.  Except during a light curve `dip' in GROJ1655-40, all 
other observed line profiles
show strong absorption features at around 7 keV. A possible interpretation is that the
continuum spectrum passes through a slab-like ionized absorption medium (accretion disk
corona) at a distance about
10$^{10}$ cm away from the original X-ray source\cite{ueda97}. The degree of ionization of the corona
increases for a higher continuum luminosity, indicating that the ionization is also caused by
the X-ray illumination. This is perhaps the first evidence of the existence of a hot corona in
a black hole binary system. It should be noted that these observed absorption features are
significantly different from the apparent emission feature observed in Cyg X-1. It is not
clear if this is due to the differences in their nature (for example, jet vs. non-jet
systems), or the inclination angle difference (70\degg\ vs 10\degg -40\degg). 

On the other hand,
the apparent emission feature in the `dip' spectrum of GROJ1655-40 should not be ignored
completely. This feature is very similar to the broadened and skewed line profiles observed
from many Seyfert I galaxies (e.g. ref. \cite{fabian95,iwasawa96,tanaka95}). The only
difference seems to be the much higher peak energy at $\sim$7.0 keV, possibly blue-shifted,
if the original peak is also the 6.4 iron $K_{\alpha}$ line. 
This is actually expected due to the much higher inclination
angle of 70\degg\ in GROJ1655-40. It is also unlikely that the whole complex line features of other
spectra (non-`dip') of the sources can be
explained completely with absorptions alone. It is possible that the non-`dip' line profiles consist of both
the absorption by the disk-corona and the iron fluorescence emission (relativistically
broadened and skewed) from the inner disk region,
which is believed to be around 1 $R_{s}$\cite{zhang-1655,zhang-cui-chen}. One possible
scenario is that during the `dip' period the emitted X-ray luminosity from the central region of the
disk becomes lower, due to perhaps the reduced mass transfer rate. The lower X-ray flux
therefore changes the properties of the disk-corona (geometry and ionization state), so that the
resonance absorption of X-ray photons by iron ions becomes negligible.

It is clear from the above discussion that rich information may be derived from iron line
diagnostics of black hole binary systems. Many of the current interpretations are, however,
quite uncertain, due primarily to the limited number of detections and the poor statistical
quality of past observations. 

\subsection*{Timing Diagnostics}

Rapid variability studies of their X-ray light curves provide essential information about many
dynamic processes in the central region of the accretion disk and in the corona. Many of the
earlier results were reviewed by van der Klis\cite{klis95}. Some recent results are reviewed
in a companion review article in the same proceedings\cite{grove-review}. Here we only review
briefly the high frequency QPOs detected from the two superluminal jet sources GROJ1655-40
at $\sim$300 Hz\cite{remillard96} and GRS1915+105 at $\sim$67 Hz\cite{morgan97}. An important 
feature in these QPOs is their {\it stable} frequency, compared to
the majority of other QPOs observed in black hole and neutron star binaries, with the
exception of the stable 34 Hz QPO in the bursting pulsar GROJ1744-28 \cite{will-1744}. 

It is thus natural to relate these QPOs to the last stable orbit of the black hole. 
Assuming that the QPO frequencies are actually the Keplerian frequencies of the last stable 
orbit, the
masses of the assumed non-spinning black holes are around 33 \msun\ and 7 \msun\ (consistent
with its dynamical mass estimate), in GRS1915+105 and GROJ1655-40, respectively. However, since the 
spin of the black hole in GROJ1655-40 is inferred
to be near its maximal rate from X-ray spectroscopic measurements as we discussed
above\cite{zhang-cui-chen}, the non-spinning black hole assumption is not likely to be valid 
and thus neither is this
Keplerian frequency interpretation. The other interpretation is to attribute these QPOs to the
$g$-mode oscillations due to the general relativity 
effects\cite{nowak-wagoner93,nowak96,perez95}. When applied to GROJ1655-40, the inferred black
hole spin rate is remarkably consistent with that obtained from independent spectroscopic
measurements\cite{zhang-cui-chen}. Taking the same approach, the implied black hole spin in
GRS1915+105 would also be near its maximal rate and the black hole mass is around 30 \msun.
In this model, however, it is not clear why the QPOs seem to be associated with the hard 
X-ray component \cite{morgan97,remillard96}. 
 
\section*{Summary}

We have reviewed the recent progress in the investigations of the multifrequency properties
and connections of galactic black hole binaries. There is now sufficient evidence for the
existence of stellar mass black holes in at least 10 systems. Their high energy and radio
light curves are well correlated and can be divided into three classes, which may be
related to their respective types of companion stars. Their X-ray spectral properties may be
related closely to the spins of the black holes, in a newly proposed unification scheme of all
types of black hole binaries. The correlation between the hard X-ray and radio flares
suggests that the radio emission and therefore the associated outflows are related to the
corona in the inner disk region. Iron line studies in the two superluminal jet sources provide
possible evidence of a hot disk-corona. Only the black holes in the jet systems
are found to be likely spinning at nearly the maximal rates, suggesting that the highly relativistic
jets are related to the spin of the black hole. This supports strongly
the previously suggestions
that the highly relativistic jets in AGNs might be related to the rapid black hole spins
\cite{blandford-znajek,blandford-levinson,livio96-iau}. Therefore the eventual 
unification of all types
of black hole systems (galactic and extra-galactic) becomes possible using the black hole spin
as one of the key parameters.

However, many of the current interpretations are
still quite uncertain, due to both of our limited theoretical understanding of the detailed
physics involved and the limitations of the current observations.
Therefore more theoretical investigations and future multifrequency observations are required.
Nevertheless, such recent progress suggests that the current
observational and theoretical studies of the galactic black hole binary systems have begun to
show the premise of using these systems as the laboratories for testing some physical
laws in the strongest gravitational field. Compared to AGNs in which similar studies can also
be carried out and in fact were started much earlier, the galactic black hole binary systems
are more advantageous because a) the dynamical
time scales in them are much shorter; b) their inner disk regions are directly observable in the X-ray
frequency range; c) the fine details of their jets are observable because they are much closer
to us; and d) the system parameters
can be measured through optical/infrared observations.

We thank Wan Chen, Ken Ebisawa, Jerry Orosz, Craig Robinson and Kinwah Wu for
kindly providing data and figures to be included in this review paper. Part of the V-band
light curve of GROJ1655-40, provided by Jerry Orosz and obtained by Charles Bailyn and various 
Yale observers and by Jeff McClintock, has not been published. SNZ also appreciates
those stimulating discussions with many of his colleagues, including Didier Barret, Sandip
Chakrabarti, Wei Cui, Wan Chen, Ken Ebisawa, Rob Fender, Jerry Fishman, Eric Grove,
Jean-Pierre Lasota, Hui Li, Edison Liang,
Mario Livio, Jeff McClintock, Ramesh Narayan, Bill
Paciesas, Ron Remillard, Craig Robinson, Marco Tavani, Lev Titarchuk, Jan van Paradijs and
Kinwah Wu, etc. 
 
\end{document}